# DEEPMIR: A Deep Neural Network for Differential Detection of Cerebral Microbleeds and IRon Deposits in MRI

Tanweer Rashid[1,2]*, Ahmed Abdulkadir[1,3], Ilya M. Nasrallah[1,4], Jeffrey B. Ware[4], Hangfan Liu[1], Pascal Spincemaille[5], J. Rafael Romero[6], R. Nick Bryan[4,7], Susan R. Heckbert[8], Mohamad Habes[1,2]*

[1]Center for Biomedical Image Computing and Analytics (CBICA), University of Pennsylvania, Philadelphia, PA, USA

[2]Neuroimage Analytics Laboratory (NAL) and the Biggs Institute Neuroimaging Core (BINC), Glenn Biggs Institute for neurodegenerative disorders, University of Texas Health Science Center at San Antonio (UTHSCSA), San Antonio, Texas, USA

[3]University Hospital of Old Age Psychiatry and Psychotherapy, University of Bern, Bern, Switzerland

[4]Department of Radiology, Hospital of University of Pennsylvania, Perelman School of Medicine of the University of Pennsylvania, Philadelphia, PA, USA

[5]Department of Radiology, Weill Cornell Medical College, New York, NY, USA

[6]Department of Neurology, School of Medicine, Boston University, Boston, MA, USA

[7]Department of Diagnostic Medicine, Dell Medical School, University of Texas at Austin, Austin, TX, USA

[8]Department of Epidemiology and Cardiovascular Health Research Unit, University of Washington, Seattle, WA, USA

Corresponding author: rashidt1@uthscsa.edu, habes@uthscsa.edu

## Abstract

Lobar cerebral microbleeds (CMBs) and localized non-hemorrhage iron deposits in the basal ganglia have been associated with brain aging, vascular disease and neurodegenerative disorders. Particularly, CMBs are small lesions and require multiple neuroimaging modalities for accurate detection. Quantitative susceptibility mapping (QSM) derived from in vivo magnetic resonance imaging (MRI) is necessary to differentiate between iron content and mineralization. We set out to develop a deep learning-based segmentation method suitable for segmenting both CMBs and iron deposits.

We included a convenience sample of 24 participants from the MESA cohort and used T2-weighted images, susceptibility weighted imaging (SWI), and QSM to segment the two types of lesions. We developed a protocol for simultaneous manual annotation of CMBs and non-hemorrhage iron deposits in the basal ganglia. This manual annotation was then used to train a deep convolution neural network (CNN). Specifically, we adapted the U-Net model with a higher number of resolution layers to be able to detect small lesions such as CMBs from standard resolution MRI. We tested different combinations of the three modalities to determine the most informative data sources for the detection tasks.

In the detection of CMBs using single class and multiclass models, we achieved an average sensitivity and precision of between 0.84-0.88 and 0.40-0.59, respectively. The same framework detected non-hemorrhage iron deposits with an average sensitivity and precision of about 0.75-0.81 and 0.62-0.75, respectively.

Our results showed that deep learning could automate the detection of small vessel disease lesions and including multimodal MR data (particularly QSM) can improve the detection of CMB and non-hemorrhage iron deposits with sensitivity and precision that is compatible with use in large-scale research studies.

## Introduction

The aging brain is subject to various irreversible changes, some driven by the aging process itself and others that are associated with various pathologies, including vascular lesions and neurodegeneration[1-4]. On magnetic resonance imaging (MRI), particularly tuned to be sensitive for differences in magnetic susceptibility, focal accumulations of iron content can be visible. This includes lesions with iron content such as cerebral microbleeds (CMBs) and non-hemorrhage iron deposits in the basal ganglia. CMBs are small hemorrhages that can occur sporadically throughout the brain[5]. CMBs have been associated with cognitive decline and dementia[6], and are considered a biomarker for small vessel diseases. The presence of lobar CMBs is also a marker for cerebral amyloid angiopathy[7-9]. Non-hemorrhage iron deposits are located in the deep structures of the brain, particularly in the basal ganglia. While an increase in iron concentration in the basal ganglia is expected in healthy aging[10], focal accumulation of iron has been associated with neurodegenerative disorders in small scale studies[11-13].

Most of our knowledge on the iron toxicity in the aging brain is limited by the fact that both CMBs and iron deposits could be difficult to distinguish from each other and from other similar lesions including calcification using conventional MRI techniques[14]. T2* gradient-recalled echo (GRE) and susceptibility-weighted imaging (SWI) are often used to clinically characterize CMB, with the latter being more sensitive for detecting CMBs[15,16]. CMBs can occur anywhere and appear as small rounded or ellipsoidal hypo-intense regions with a diameter of ten millimeters or less[7,14,17]. Non-hemorrhage iron deposits in the basal ganglia have irregular shapes and could be larger than CMBs[14]. Because hypo-intensities in SWI are not specific to CMBs and non-hemorrhage iron deposits, images with other tissue contrasts are required in order to identify other lesion types that can have similar low susceptibility signal on SWI, such as calcification[5,18,19]. The specificity for CMB detection can be increased by post-processing SWI-magnitude and phase data to derive quantitative susceptibility maps (QSM)[20,21]. In QSM paramagnetic tissue appears different from diamagnetic materials, and therefore this contrast is particularly useful for distinguishing non-hemorrhage iron deposits from calcifications[22,23]. While previous efforts have been made to automate the detection of microbleeds, all previous work neglected the detection of non-hemorrhage iron deposits in such automated framework[24-30].

No work has been published to date on segmenting iron deposits in the brain using QSM with either a semi- or a fully automatic method. The advances made in MRI technology with QSM for iron content recognition are gaining more attention as cohort-based studies such as The Multi-Ethnic Study of Atherosclerosis (MESA)[31-33] include QSM in their imaging protocol, and thus exploit its advantages in delivering specific insights on iron toxicity in the aging brain. The focus in MESA is utilizing non-invasive methods to investigate common risk factors, preclinical disease states and manifest diseases using a standardized imaging protocol, which is applied to all participants[34]. On one hand, this is providing a unique opportunity to study widely ignored lesions such as iron deposits in vivo using MRI but on the other hand, this comes with additional challenges as such cohorts naturally include largely cognitively normal participants with a low lesion load, resulting in a very challenging task to automate.

In order to tackle the challenges inherent in the detection of these lesions, we developed a robust and fully automated deep learning-based method to detect CMBs and non-hemorrhage iron deposits in a cohort without extensive apparent brain tissue damage and having a low load of CMBs and non-hemorrhage basal ganglia iron deposits. We experimented with both single class and multiclass segmentation models using multiple MR sequences. Our experiments show that using multi-sequence MRI (especially QSM) improves the overall accuracy of detection. The main contributions of this study include the following:

1. We tackled the challenging problem of simultaneously detecting CMB and non-hemorrhage iron deposits. To our knowledge, this is one of the first reports to detect both types of lesions simultaneously. Often iron accumulation in the brain has been understudied due to the lack of appropriate techniques for detecting them in vivo in large-scale epidemiological studies;
2. We found out the most suitable pulse sequence combination to automate the detection tasks by exploiting imaging information jointly;

3. We developed an effective and flexible neural network model that is specially tailored to the differential detection task. The proposed model can be easily adapted to segment additional lesions;
4. We achieved highly competitive detection performance on real-life data, demonstrating the effectiveness of the proposed approach in practical applications.
5. We also provide access to our source code and a few trained models via the GitHub link https://github.com/NAL-UTHSCSA/CMB_NHID_Segmentation

## Results

We performed leave-one-out cross-validated evaluations for both single class and multiclass segmentation experiments using the 24 participants listed in Supplementary Table 2.1. Panel A in Figure 1 shows an example of the automated segmentation of a CMB (indicated by the red arrow). Panel B in Figure 1 shows the segmentation of the focal iron deposits in the basal ganglia. In this figure, the model correctly segmented the iron deposit lesions (indicated by the green arrow) while rejecting an instance of calcification (indicated by the yellow arrow). The results of these experiments are reported in Table 1 and Table 2. Pearson's correlation and Bland-Altman mean difference and confidence intervals for single class and multiclass experiments are reported in Table 1 and Table 2 respectively, for the experiments with 24 participants. Overall, our experiments show that incorporating QSM in model training can increase the overall accuracy of CMB and iron deposit detection.

In the case of segmenting CMBs, the best performance in terms of average magnitude accuracy is seen with the model trained with SWI and QSM in both single class and multiclass experiments. The correlation coefficient between the prediction and ground truth was also highest (r=0.97 and r=0.99, for single class and multiclass results, respectively) when QSM was included in the training. For non-hemorrhage iron deposits, the single class model trained with all three modalities had the highest average magnitude accuracy and the multiclass model trained with SWI and QSM had the highest average magnitude accuracy. The correlation coefficient was also highest for models that included QSM for training (r=0.92 and r=0.91 for single class and multiclass results, respectively). Figure 2 shows a joint scatterplot of the single class experimental results and Figure 3 shows a joint scatterplot of the multiclass experiments.

In our dataset, we identified as an outlier a single individual with exceptionally many CMBs. A comparative analysis was done by removing this outlier from the dataset and repeating a similar cross-validated evaluation by retraining both single class and multiclass models. The results are detailed in Supplementary Table 3.1 and Supplementary Table 3.2 in Section 3 of the supplementary materials. With the exception of multiclass CMBs, we note that the best result in terms of magnitude accuracy is seen when the model training includes QSM. This is also reflected by the correlation coefficients. Models which included QSM showed a higher correlation between the number of predicted lesions and reference annotation. For CMBs, the correlation r=0.51 and r=0.69, for single class and multiclass results, respectively. For iron deposits, the correlation r=0.94 and r=0.97 for single class and multiclass results, respectively.

We have conducted another experiment where we modified the proposed DEEPMIR architecture by removing one layer. The results of this experiment are reported in Supplementary Table 4.2 in Section 4 of the supplementary materials. For detecting CMBs, the model trained with SWI, QSM and T2w has the highest magnitude accuracy. For detecting iron deposits, the model trained with SWI and QSM showed the highest magnitude accuracy but having a lower correlation coefficient (r=0.75), while the model trained with SWI, QSM and T2w reports similar magnitude accuracy and higher correlation coefficient (r=0.93).

An additional leave-one-out cross-validated evaluation was done for the 24 participants using an implementation of the original U-Net[35]. The results of this experiment are reported in Supplementary Table 4.1 in Section 4 of the supplementary materials. In these experiments, we note that the models trained with SWI, QSM and T2w had the best performance in terms of magnitude accuracy for both CMBs and iron deposits. However, in terms of the correlation coefficient, we note that the model trained with SWI and QSM had the highest correlation (r=0.98 and r=0.93, for CMBs and iron deposits, respectively).

We investigated the performance of the proposed DEEPMIR architecture for the simultaneous differentiation and labeling of both CMB and iron deposit labels against the performance of the original U-Net and a modified DEEPMIR architecture (having the same number of resolution layers as the original U-Net). We note that the proposed DEEPMIR model with 6 resolution layers has better overall sensitivity for detecting small lesions such as CMBs. Supplementary Figure 12 and Supplementary Figure 13 show examples of small lesions that the original U-Net and the modified DEEPMIR models were unable to detect, compared to the accurate detection by the proposed DEEPMIR architecture.

## Discussion

We developed a deep learning framework for simultaneous segmentation of cerebral microbleeds and non-hemorrhage iron deposits using multi-modal MRI. To date, previously published methods for automated or semi-automated CMB detection have ignored iron deposits. In this study, we consider the iron deposit in the basal ganglia seen as hypo-intense lesions on SWI and confirmed by QSM to be iron-specific rather than mineralization. Those lesions may typically be labeled as possible or uncertain microbleeds on MARS[18] and BOMBS[19] mainly because of the limitation that T2* and SWI cannot differentiate iron content from mineralization. We overcome this limitation by including QSM in our study, which has shown to improve the overall accuracy for automated detection. To our knowledge, there are no studies that attempted to segment these focal iron deposits using SWI and/or QSM automatically. Our deep learning-based segmentation method presented here is filling in this gap. We have undertaken several experiments using both single class and multiclass models with different combinations of the available MR pulse sequences. We noted that the models which included QSM in training consistently performed better and the resulting predictions had statistically high correlations when compared to the reference annotation.

Our approach has several advantages over the current state-of-the-art methods for CMB detection. First, by using deep learning our model is capable of learning and generalizing features rather than rely on feature vectors derived with conventional image processing algorithms[28-30], Fourier shape descriptors[36] or probabilistic models[27]. Second, we employ end-to-end learning by using a single model (or network). Previously published methods that used deep learning employed multiple stages consisting of (a) a candidate generation stage which use either conventional image processing methods[24,26] or an initial (and separate) deep learning-based model[25] for identifying possible CMBs, and (b) a false positive reduction stage in the form of a CNN-based network[24-26]. Our single-stage design allows for greater flexibility, for example in retraining with different or larger data sets, adding additional class labels, or using different modalities, while achieving sensitivity and precision comparable to published results. Third, we trained with different sets of input imaging modalities. Combinations of imaging modalities allowed our models to reject mimics such as calcifications without explicit provisions (as shown in Figure 1, Panel B). Supplementary Figure 9 in the supplementary materials (Supplementary Section 5) shows an example of mineralization being segmented as iron deposits when the model was trained with only SWI. The models in publications[25,26] used SWI only and therefore may not be capable of recognizing and rejecting mimics. The method in publication[24] utilizes SWI-phase and magnitude images along with QSM, but did not consider iron deposits in the basal ganglia. Fourth, we experimented with a reduced number of layers (5 instead of 6 spatial resolution layers) and noted that having more layers can improve the overall results for detecting small lesions such as CMBs.

Our framework has achieved an excellent sensitivity of 89%. However, other studies[24-26] have reported higher precision in detecting CMBs in their samples. We would like to note that it is impossible to directly compare reported numbers from various machine learning models, due to differences in populations included, study settings and imaging and scanner charactherisitcs[37]. Of particular importance is the fact that our sample was drawn from a relatively healthy population without significant brain trauma, injuries, or pre-existing neuro-pathologies whereas the studies in publications[24-26] had hundreds if not thousands of CMB lesions related to or caused by radiation therapy, stroke and traumatic brain injury.

One of the major challenges was the small size of the lesions and their potential presence throughout the brain. The average size of four voxels (or 6 mm$^3$) per CMB together with the generally low lesion burden of the study participants resulted in including only two CMB lesions/4 voxels on average per participant, resulting in a higher weight of a single lesion or error in the evaluation. In other words, missing a single lesion would result in a drop of sensitivity from one to 0.5 and a single false positive for a given participant would result in a drop of that participant's precision from one to 0.5 or 0.66. Similarly, a small number of false positives, in absolute terms, can lower the average precision substantially. In general, our models over-segmented the data in terms of detecting more CMBs than were actually present (Supplementary Figures 10 and 11 in the supplementary materials show examples of false positive CMBs). In all experiments using the aforementioned combinations of available imaging modalities, most of the lesions were detected and the average sensitivity was consistently above 0.75.

Notably, the sample used to train the model was a convenience sample from participants of the MESA study without particular clinical profile and without apparent brain disorders such as dementia, depression, or traumatic brain injury. Given the low number of lesions on average, our method achieved sensitivities that are comparable to state-of-the-art CMB segmentation/detection methods trained with large datasets. We expect that including more samples with more lesions would improve the precision. In general, most studies incorporating automated methods for large-scale abnormality detection or brain region segmentation incorporate a segmentation quality control step that could result in corrections or exclusions[1,38,39]. Thanks to the flexibility of our method, it is straightforward to increase the sample size.

In clinical terms, a larger number of CMBs is more likely to be clinically relevant. The proposed DEEPMIR method was trained and evaluated on a relatively small population and outputs the number of lesions and lesion segmentation maps for each participant. The next step would be to rigorously test and evaluate the proposed model on a larger sample size to ensure viable sensitivity, precision and overall accuracy, before applying it to a large cohort to determine the prevalence of lesions in the population. An adequately trained model can be used as a screening tool to flag participants with a high lesion load. DEEPMIR can also be used to generate an initial segmentation of lesions to accelerate manual annotation.

QSM is a good, non-invasive technique to distinguish between iron content and mineralization in the brain and showed a great advantage in improving the overall accuracy of CMB and iron deposit detection in the current study. While QSM is being recognized and is being integrated in more population-based studies, large studies with QSM data acquisition such as MESA is still ongoing. This left us with a relatively small number of imaging data used for training. For our experiment, we had a ratio of validation to training data (25:75), which showed to be reasonable to ensure that a maximal amount of the available data is used in model training, while at the same time a sufficient amount is reserved for within-training validation. The use of similar sample sizes for training and evaluation is not unprecedented in such small lesion detection[27,29,40,41]. One limitation of using such a small sample size is a reduction in study statistical power. For our experiments, we noted that none of the multiple comparisons were statistically significant, and this could likely be due to the small sample size. Finally, the limited access to QSM from other studies left us to perform cross-validation[37] with samples from only the MESA AFib cohort for evaluating our model. We were therefore not able to test the generalizability of our model with images generated in other studies with different parameters and characteristics. This line of work should be considered in future research efforts, ultimately building machine learning models and benefiting from pooling imaging data from multiple cohort-based studies[42].

We have presented a framework for the automated detection of cerebral microbleeds and non-hemorrhage iron deposits in the basal ganglia. While SWI remains the preferred modality of choice for CMB detection, few studies have leveraged QSM as an additional source of information to improve overall detection accuracy, and to date, there have been no attempts to include iron deposits in the basal ganglia as an item of interest. We have utilized QSM in this study to confirm that these focal lesions in the basal ganglia are in fact iron depositions, rather than mineralization such as calcifications. Our deep learning neural network model is flexible and at the same time scalable to include additional modalities and/or class labels while maintaining comparably high sensitivity and precision. We aim in our future work to automatically detect other small vessel disease lesions in our framework such as enlarged perivascular spaces. We also aim to investigate possible advantages of expanding our network to a three-dimensional variant.

## Methods

### MRI Acquisition and Pre-Processing

The MESA Exam 6 Atrial Fibrillation (AFib) Ancillary Study's[34] brain MRI protocol included T1-weighted (T1w) and T2-weighted (T2w) sequences, and a susceptibility weighted imaging (SWI) sequence with 4 different, equally spaced echo times. SWI is a high-resolution, 3D imaging sequence where the image contrast is enhanced by combining magnitude and phase image data[43,44]. The scans were acquired at 6 sites using the same acquisition parameters. All scans were performed on Siemens MR scanners (2 Skyra with a 20-channel head coil and 4 Prisma Fit with a 32-channel head coil) at a static magnetic field strength of 3 Tesla and identical imaging sequence parameters, as shown in Supplementary Table 1.1 in Section 1 of the supplementary materials.

Multiple SWI phase and magnitude images were acquired with varying echo times (Supplementary Table 1.1 in the supplementary materials). SWI data were generated following the method of Haacke et al.[43,45]. A homodyne high-pass filter with k-space window size of 64 x 64 was applied to the raw phase image to generate the negative phase mask (with values between 0 and 1). The phase mask was then raised to power 4 and multiplied with the magnitude image to generate the SWI. For creation of the reference annotation and subsequent deep learning-based inferencing, only the SWI image with the shortest echo time (TE=7.5 ms) was used because longer echo times have more noise due to increasingly pronounced blooming effects near the sinus cavity and cerebellum. In addition, SWI with longer echo times are also more prone to showing false positive CMBs, especially when veins are perpendicular to the imaging plane. Section 2 in the supplementary materials discuss this issue in more detail.

The T1w and T2w images underwent N4 bias correction[46] with default parameters using the implementation in the Advanced Normalization Tools (ANTs) (http://stnava.github.io/ANTs) suite and were rigidly registered to the participants' SWI image using FSL's FLIRT[47-49] (https://fsl.fmrib.ox.ac.uk). Anatomical parcellation and brain masks were generated with a multi-atlas segmentation method using the bias-corrected T1w images[50]. These brain masks were used in the generation of the QSM images. QSM maps were generated using the entire multi-echo SWI dataset using the Morphology Enabled Dipole Inversion (MEDI) method[21,51] implemented in MATLAB (http://weill.cornell.edu/mri/pages/qsm.html). Briefly, background field removal is done using the Projection onto Dipole Fields (PDF) method[52], followed by region-growing based spatial unwrapping with non-linear fitting[53] to reduce errors, and finally the susceptibility map is calculated using the Morphology enabled dipole inversion with zero reference using CSF (MEDI+0) method[54].

**Manual Annotation**

Manual annotation was performed according to a protocol developed with the focus on highly specific differential detection of CMBs and non-hemorrhage iron deposits based on multiple modalities including QSM. The detailed protocol is described in Section 2 in the supplementary materials, and a flowchart of the manual annotation process is shown in Supplementary Figure 4 in the supplementary materials. Panel A in Figure 4 shows an example of a CMB in the thalamus and non-hemorrhage iron deposits in the interior section of the globus pallidus on SWI (for TE=7.5 ms and 22.5 ms), QSM and T2w MRI, and Panel B shows the expert segmentation of the lesions based on the annotation protocol. Panel C shows an example of a larger CMB located in the occipital lobe and Panel D shows its respective expert segmentation.

**Study Participants**

We included imaging data from participants in the MESA Exam 6 Atrial Fibrillation Ancillary Study[31-33]. This study was approved by the Institutional Review Boards at the MESA Coordinating Center and at each participating institution. Written informed consent was obtained by all participants. All participant data collection was performed in accordance with relevant guidelines and regulations.

A subset of the MESA cohort participated in an ancillary study of cardiac arrhythmias and brain imaging during the 2016-2018 exam (Exam 6)[34]. From 1061 participants who underwent MR brain scans, we selected a convenience sample of 34 scans based on visual identification of possible CMBs by two experienced readers (IMN and TR). These 34 participants are not representative of the MESA cohort in terms of prevalence of CMBs and non-hemorrhage iron deposits, and additional participants in the MESA cohort likely have CMBs and/or non-hemorrhage iron deposits. A total of 10 participants' scans were excluded due to poor image quality ($n$=4) and the presence of distortions/artifacts or motion-related effects ($n$=6). The demographics summary and lesion loads for the 24 included participants are presented in Supplementary Table 2.1. Of these 24 participants, there were 13 males and 11 females with age range 65-94 years. Based on the expert annotation of these 24 participants, 4 participants had no microbleeds, 13 participants had 1 or 2 microbleeds (with an average size of 10.85 mm$^3$), 6 participants had between 3 and 8 microbleeds (with an average size of 10.21 mm$^3$) and 1 participant had more than 100 microbleeds (with an average size of 4.76 mm$^3$). In certain circumstances, the participant with more than 100 microbleeds may be considered an outlier in terms of the number of CMBs. An examination of this is presented in Section 3 of the supplementary materials. Of the 24 participants, 5 participants did not have any voxels labeled as non-hemorrhage iron deposits and the remaining had between 2 (each having a single voxel or 1.5 mm$^3$) and 13 lesions (one participant had 4 non-hemorrhage iron deposit lesions with a total of 326 voxels or 489 mm$^3$) labeled as non-hemorrhage iron deposits in the basal ganglia.

The distribution of CMBs and iron deposits pooled over all participants is illustrated in Supplementary Figure 5. The average size (± SEM, or standard error of the mean) of CMB lesions in this sample was 6.27 ± 0.51 mm$^3$ (4.18 ± 0.34 voxels). Among the 20 participants with CMB, 70% ($n$ =14) had two or fewer CMBs, 25% ($n$ = 5) had between three and eight CMBs, and the remaining participant had 120 CMBs. The average size of non-hemorrhage iron deposit labels (± SEM) was 26.15 ± 4.76 mm$^3$ (17.43 ± 3.17 voxels). Approximately 21% ($n$ = 5) had no discernable basal ganglia non-hemorrhage iron deposits and half ($n$ = 12) had fewer than 100 voxels (150 mm$^3$) labeled as non-hemorrhage iron deposits. The remaining 29% ($n$ = 7) had more than 100 voxels labeled as non-hemorrhage iron deposits.

**Method Overview for Automated Processing**

We developed a deep learning framework for automatic segmentation of CMBs and non-hemorrhage iron deposits based on the U-Net[35,55], a widely used deep learning architecture for image segmentation. Our architecture, however, employed padded instead of unpadded convolutions and operated on six instead of five spatial resolutions, and was used for both single class and multiclass segmentation experiments. The larger number of resolution layers enabled the model to detect small CMBs. A detailed description of our implementation is presented in the following sections. The overall system pipeline is shown in Figure 5. After the initial step of co-registration, the MR volumes were preprocessed to have zero mean and unit variance, as detailed in Section 4.6. The normalized MR volumes were then sliced along the *z*-axis (axial slices) and edge-padded to obtain 2D slices with 256x256 voxels. We evaluated the performance using leave-one-out cross-validation for the 24 participants listed in Supplementary Table 2.1 to ensure generalization of results. In each fold, a single participant's data was kept separate for testing (test dataset), and the MR data and labels from the remaining 23 participants were randomly split into training dataset (75%, consisting of 17 participants) and validation datasets (25%, consisting of 6 participants). Both training and validation datasets were augmented to improve the robustness of the deep learning models (for more details on data augmentation see Section 4.7). The training dataset was used to train the model for a single epoch, after which the validation dataset was used to compute a commonly used evaluation metric known as intersection-over-union (IoU) which quantifies the amount of overlap between the predicted and ground truth segmentations. Each model was trained for a maximum of 30 epochs, and the best model was determined as the model with the maximum IoU. This best model was then used to predict the labels of the test dataset. The set of predictions used for evaluating model performance thus consisted of 24 segmentation masks that were predicted with 24 different models with no overlap between training, validation and testing datasets. These cross-validated evaluations were done for both single class and multiclass experiments. For all experiments, four permutations of MR modalities were considered: (1) SWI only, (2) SWI and QSM, (3) SWI and T2w, and (4) SWI, QSM and T2w.

For single class experiments, separate models were trained and evaluated for (1) CMBs only and (2) non-hemorrhage iron deposits only. For multiclass experiments, both CMBs and iron deposits had separate labels and were segmented simultaneously. For multiclass segmentations, a larger number of augmentations were used than for single class segmentations.

**2D U-Net with Padded Convolutions**

Our lesion prediction models are based on the U-Net[35]. Both single and multiclass models consist of an analysis path (down-sampling operations) with five stages of convolution blocks and pooling, followed by a five synthesis path (up-sampling) with five stages of up-convolutions, plus a convolutional block. Each downsampling block consists of two layers of a 2D padded convolution layer having kernel size of 3x3 and stride of 1x1, followed by Batch Normalization and ReLU activation. The downsampling block ends with a 2x2 max pooling layer which reduces the resolution feature map by half in every spatial direction. The central block consists of two instances of padded 2D convolution with kernel size 3x3 and stride 1x1, followed by Batch Normalization and ReLU activation. Each upsampling block passes its input data through a 2D transpose convolution with kernel size of 2x2 and stride 2x2 in order to double the size of the feature map. This doubled feature map is then concatenated with the feature map (same size) of the corresponding analysis stage (i.e. the feature map before max pooling layer), followed by two instances of a padded 2D convolution layer having kernel size 3x3 and stride 1x1, followed by Batch Normalization and ReLU activation. Due to the use of padded convolutions throughout the model, the input and output image sizes are the same (256x256). The smallest downsampled image size is 8x8 in the central convolution block.

In the case of the single class prediction model, the output of the final upsampling stage passes through a 2D convolution layer with kernel size 1x1, stride 1x1 and Sigmoid activation function. For the multiclass prediction model, the output of the final upsampling block is passed through a 2D convolution layer with kernel size 1x1, stride 1x1 and ReLU activation function, and then through a SoftMax layer to generate class probabilities. The model architecture is depicted in Figure 6. We employed random translations, random rotation, and flipping along the left-right axis during training. The network was trained with the cross-entropy loss.

**Image Preprocessing**

Each input image was normalized to have zero mean and unit variance. For QSM images, an additional prior step truncated the overall intensity such that the voxel value ($V_{QSM}$) was within the range $[-k * \sigma_{QSM} \leq V_{QSM} \leq k * \sigma_{QSM}]$, where k = 5 and $\sigma_{QSM}$ is the standard deviation for the QSM image. This step is necessary because QSM images contain high-intensity noise (especially around the boundary of the brain and the region proximate to the sinus cavity) which may de-emphasize the intensity of the rest of the brain.

**Data Augmentation**

To improve the robustness of the deep learning network and include more training data we enriched the training and validation datasets with augmentation. Axial slices containing CMBs and iron deposits are, for the most part, few compared to the remaining slices in a given brain volume. This type of class imbalance may bias the training process. To address this, data augmentation was performed on slices selectively instead of all slices, inspired from the concept of random over-sampling (ROS) and random under-sampling (RUS) [56]. First, all slices containing the labels of interest (i.e. CMBs and/or iron deposits) are augmented. Then a number of the remaining slices are randomly selected and augmented in the same manner until the total number of slices containing the labels of interest and the total number of slices that do not contain any labels of interest is similar.

Data augmentation consisted of geometric transforms such as translations, rotations and image mirroring. In each experiment, the axial SWI slice (along with the corresponding axial QSM and T2w slices) and corresponding axial reference annotation slice were augmented. For translations, a set of two random integers *tx* and *ty* (representing the amount of shift per axis) were generated within the range [-45, 45] and used to translate the image slice(s) and the corresponding slice of the reference annotation. This range was chosen empirically so that most of the brain would be visible in the translated image. A total of 10 random integers per axis were generated for multiclass experiments.

For rotations, a set of random integers *d* (representing the rotation in degree) were generated within the range [1, 60], and the image slice(s) and the slices with reference annotations were rotated using both +*d* and -*d*. The regions of the crops that were located outside the image matrix were padded with edge values. A total of 16 random integers were used for multiclass experiments.

## Evaluation of Performance

In single class models, the segmentation output map was in the range [0, 1]. Segmentations were accepted or rejected by applying a threshold value of 0.5 to the output map. In multiclass models, the model output was passed through a SoftMax function and segmentation labels were determined based on the class having the highest probability.

We evaluated the performance in terms of the rate of detected/missed CMBs and non-hemorrhage iron deposit lesions. For each participant, the number of true positives (TP), false positives (FP) and false negatives (FN) were counted. A connected-component filter with 3D connectivity was applied to both the predicted segmentation and the reference segmentation in order to identify clusters of voxels. The centroid of the lesion in both the predicted segmentation and reference annotation was computed. TP, FP and FN were determined on whether the Euclidean distance between the centroid of each predicted lesion and a reference lesion was below a specified tolerance. Since CMBs are generally assumed to be relatively small in size, a tolerance of 3 was used for evaluating CMBs, and a tolerance of 5 was used for evaluating non-hemorrhage iron deposits since iron deposits have a larger size and more dispersed pattern than CMBs which are spherical. The sensitivity $S$ (or true positive rate) was computed as the ratio of TP and number of lesions in the ground truth (TP + FN) for each participant:

$$S = \frac{TP}{TP+FN} \qquad (1)$$

The precision (or positive predictive value) $P$ was computed as the ratio of TP and the number of lesions in the predicted mask:

$$P = \frac{TP}{TP+FP} \qquad (2)$$

When the true negative (TN) is available, the typical measure of performance is the overall accuracy, determined by

$$ACC = \frac{TP + TN}{TP + TN + FP + FN} \tag{3}$$

To evaluate the performance of each model, we report the average sensitivity across all participants and average precision across all participants, as well as a combined metric (magnitude accuracy) computed as $\sqrt{\bar{S}^2 + \bar{P}^2}$, where $\bar{S}$ and $\bar{P}$ are the average sensitivity and precision respectively.

**Statistical Analysis**

Due to the small sample size and potentially non-uniform distribution of the models' sensitivity, precision and magnitude accuracy, we utilized the non-parametric two-tailed Wilcoxon signed rank test[57] to check for any difference between the performance of the various models. In all experimental evaluations, the model trained with only SWI was considered as the baseline model for comparison. Statistical significance was considered at a $p < 0.05$. Correlation (Pearson's) between the prediction and reference annotation is also calculated. For CMBs, the correlation was calculated using the number of lesions, and for non-hemorrhage iron deposits, the volume was used. All statistical analyses were performed in MATLAB R2017b.

## Acknowledgements

This research was supported by contracts 75N92020D00001, HHSN268201500003I, N01-HC-95159, 75N92020D00005, N01-HC-95160, 75N92020D00002, N01-HC-95161, 75N92020D00003, N01-HC-95162, 75N92020D00006, N01-HC-95163, 75N92020D00004, N01-HC-95164, 75N92020D00007, N01-HC-95165, N01-HC-95166, N01-HC-95167, N01-HC-95168 and N01-HC-95169 and grant HL127659 from the National Heart, Lung, and Blood Institute, and by grants UL1-TR-000040, UL1-TR-001079, and UL1-TR-001420 from the National Center for Advancing Translational Sciences (NCATS). The content is solely the responsibility of the authors and does not necessarily represent the official views of the National Institutes of Health. The authors thank the other investigators, the staff, and the participants of the MESA study for their valuable contributions. A full list of participating MESA investigators and institutions can be found at http://www.mesa-nhlbi.org.

## Author Contributions Statement

T.R. wrote the manuscript and was responsible for code development and statistical analysis. A.A. contributed to code development, writing and editing. H.L. contributed to writing and editing. P.S. contributed code and critical review. I.M.N., J.B.W., J.R.R., S.R.H. provided critical review of the manuscript. M.H. and S.R.H. acquired funding for the study. M.H. contributed to study planning, experimental design, supervision, writing and critical review of the manuscript. All authors contributed critical review and approval.

## Additional Information

### Competing Interests

The authors declare that they have no competing interests.

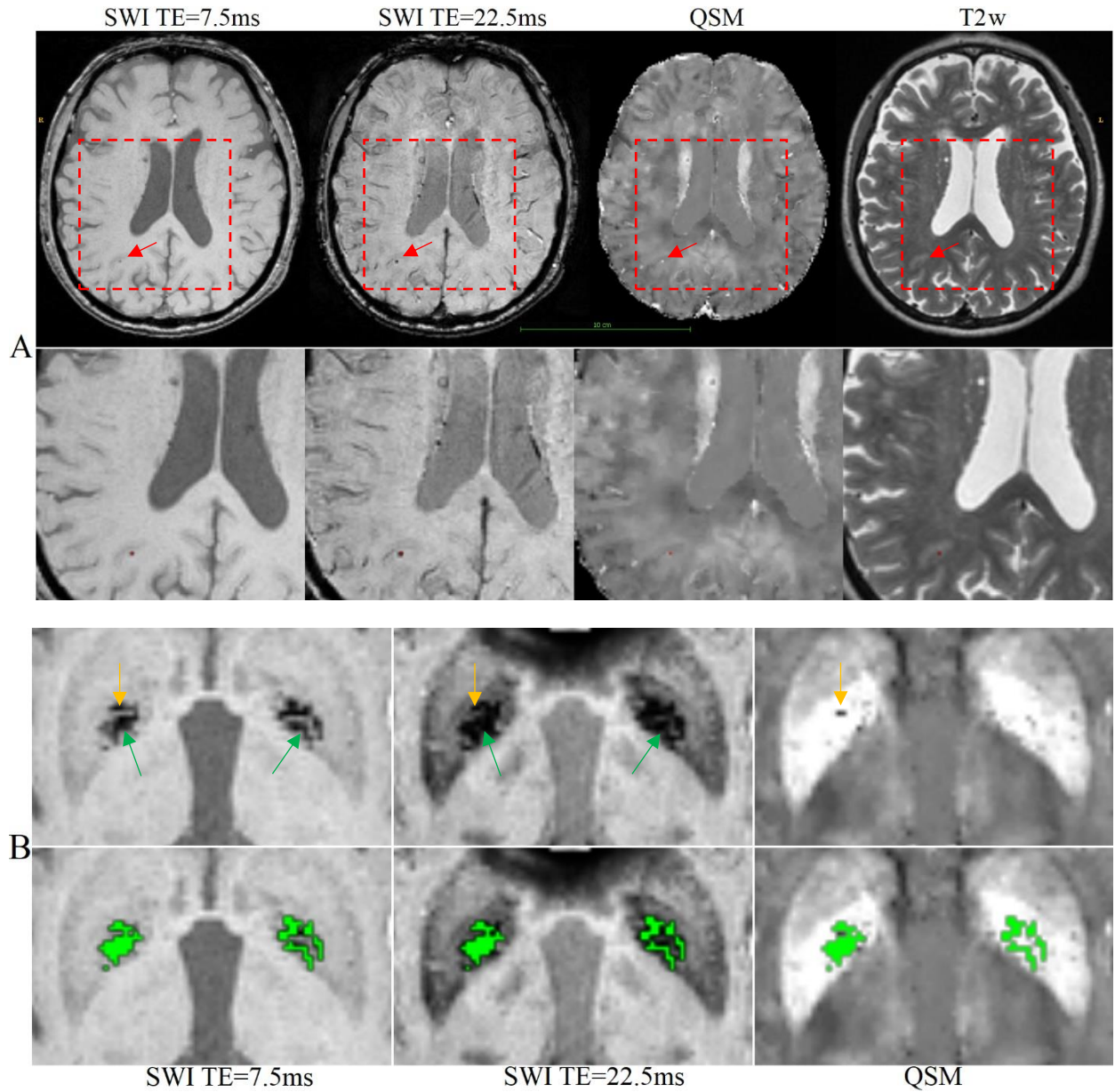

**Figure 1.** Panel A: Segmentations of CMBs by a model trained with SWI, QSM and T2w. (Top) An example of the correct segmentation of a small microbleed (red arrow). (Bottom) Magnified view of microbleed with segmentation mask (single red pixel). Panel B: An example of QSM being used to distinguish iron deposits from calcifications in the basal ganglia. (Top row) The SWI for TE=7.5ms, SWI for TE=22.5ms and QSM of the basal ganglia. The yellow arrow points to hypo-intense voxels which are likely calcifications and the green arrow points to basal ganglia iron deposits. (Bottom row) The segmentation mask (green labels) of the iron deposits. For both Panel A and B, the segmentations were generated by the multiclass model trained with SWI, QSM and T2w.

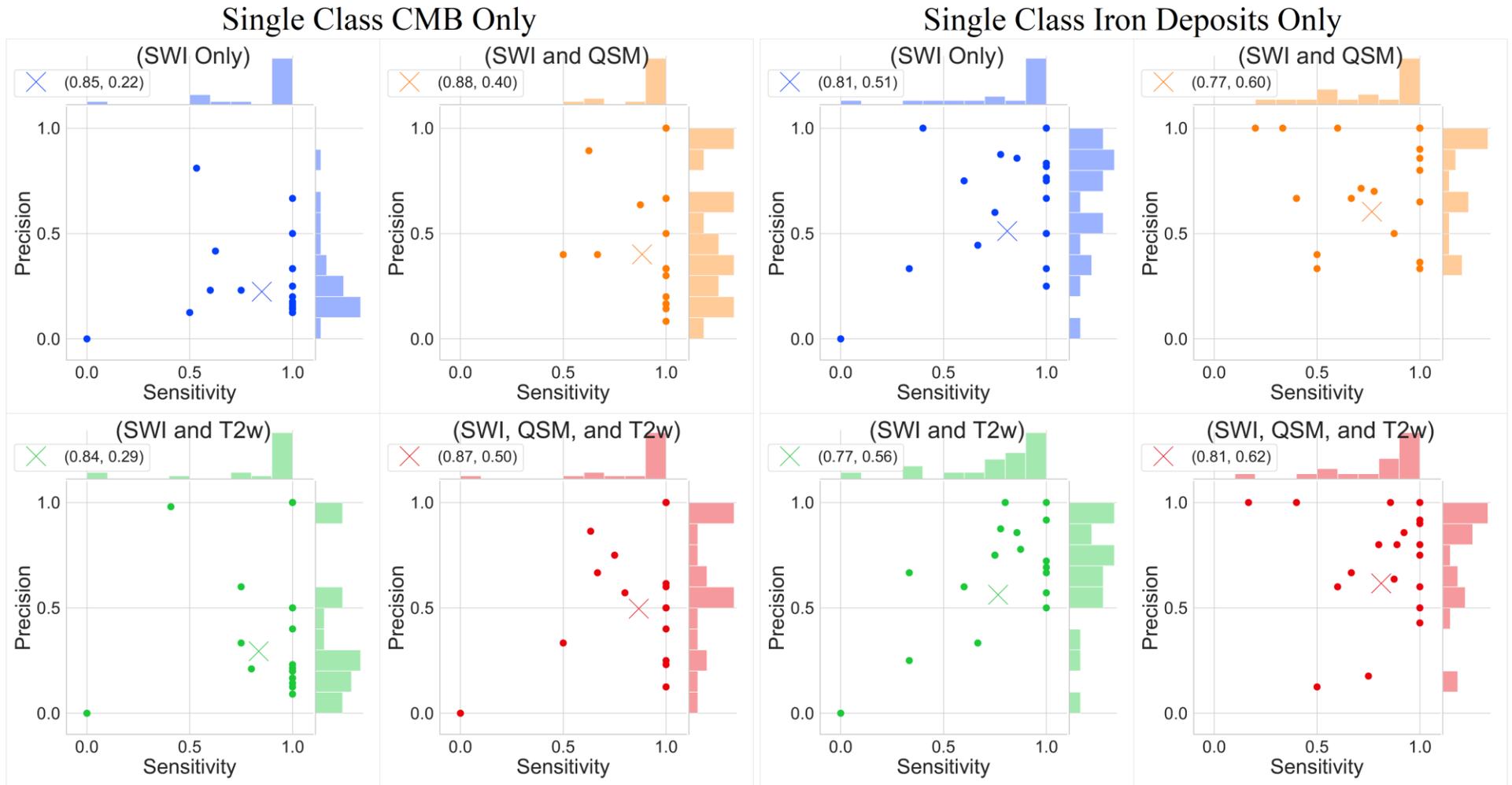

**Figure 2.** Joint scatterplots of the sensitivity vs precision of all single class experiments predicting CMBs and non-hemorrhage iron deposits. (Left) all CMB only experiments and (Right) all iron deposits only experiments. In each subplot, the round points indicate the individual participants' sensitivity and precision evaluated with leave-one-out cross-validation, and the X indicates the mean sensitivity and precision. The legend at the upper left corner of each subplot shows the coordinates of X. In each subplot, histograms of the sensitivity and precision are displayed along the upper and right axes.

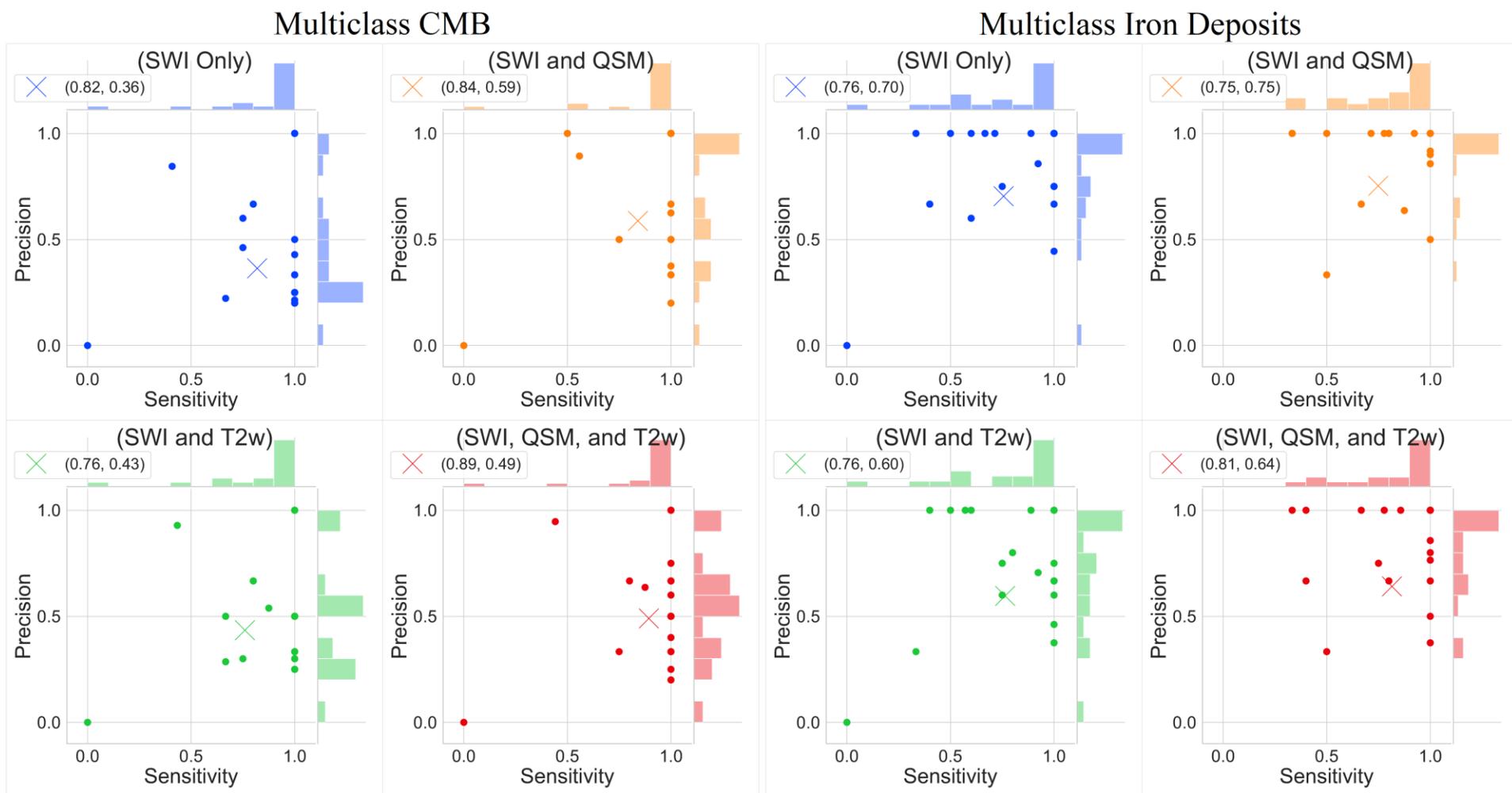

**Figure 3.** Joint scatterplots of the sensitivity vs precision of all multiclass experiments predicting CMBs and non-hemorrhage iron deposits. (Left) all evaluations for CMBs and (Right) all evaluations for iron deposits. In each subplot, the round points indicate the individual participants' sensitivity and precision evaluated with leave-one-out cross-validation, and the X indicates the mean sensitivity and precision. The legend at the upper left corner of each subplot shows the coordinates of X. In each subplot, histograms of the sensitivity and precision are displayed along the upper and right axes.

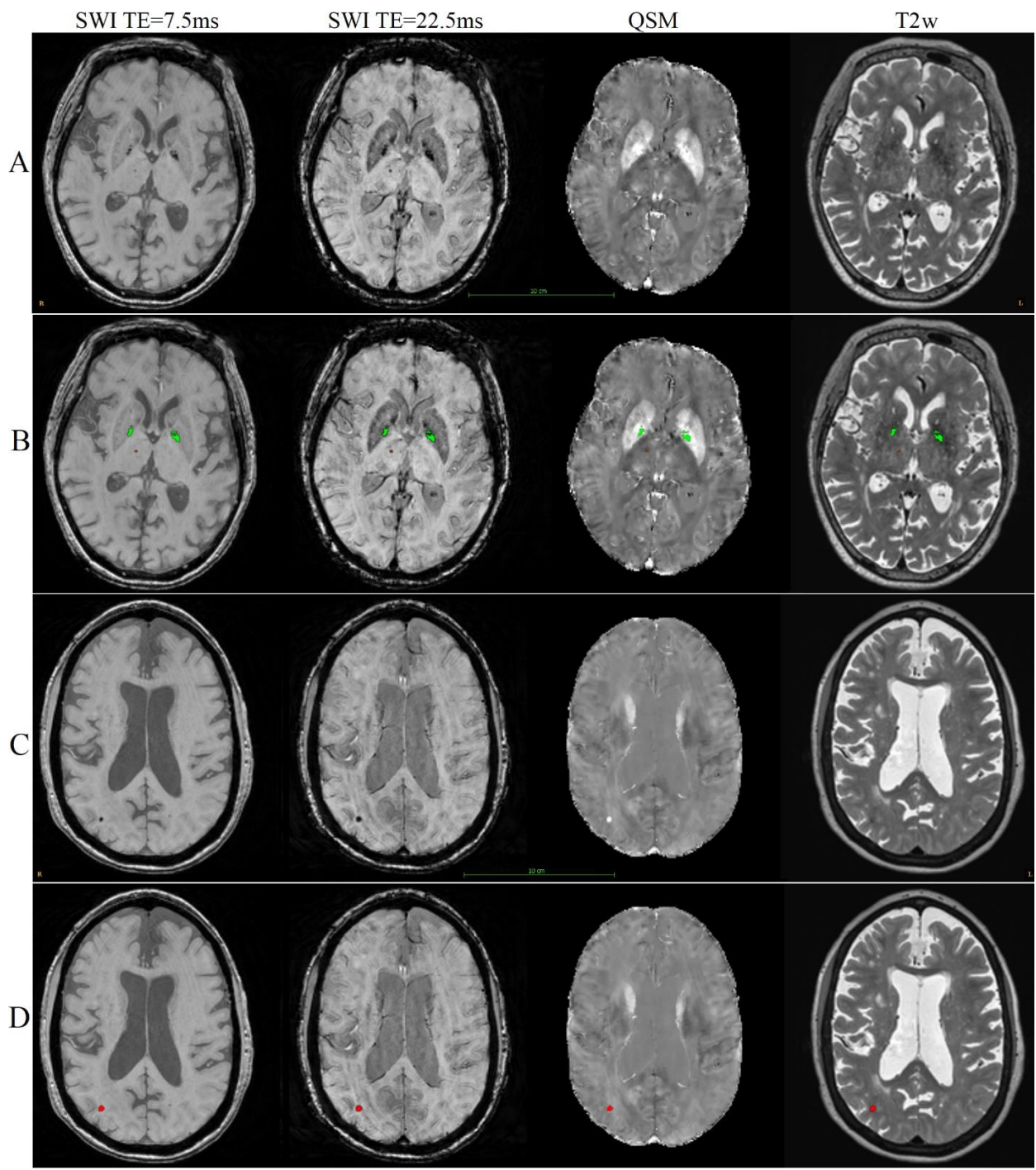

**Figure 4.** This figure shows examples of cerebral microbleeds and basal ganglia iron deposition in SWI for TE=7.5ms (left column), SWI for TE=22.5ms (middle left column), QSM (middle right column) and T2w (right column). Panels A and C show the lesions in two different brains, and Panels B and D show the corresponding human expert labeling of the CMBs (red) and iron deposits (green).

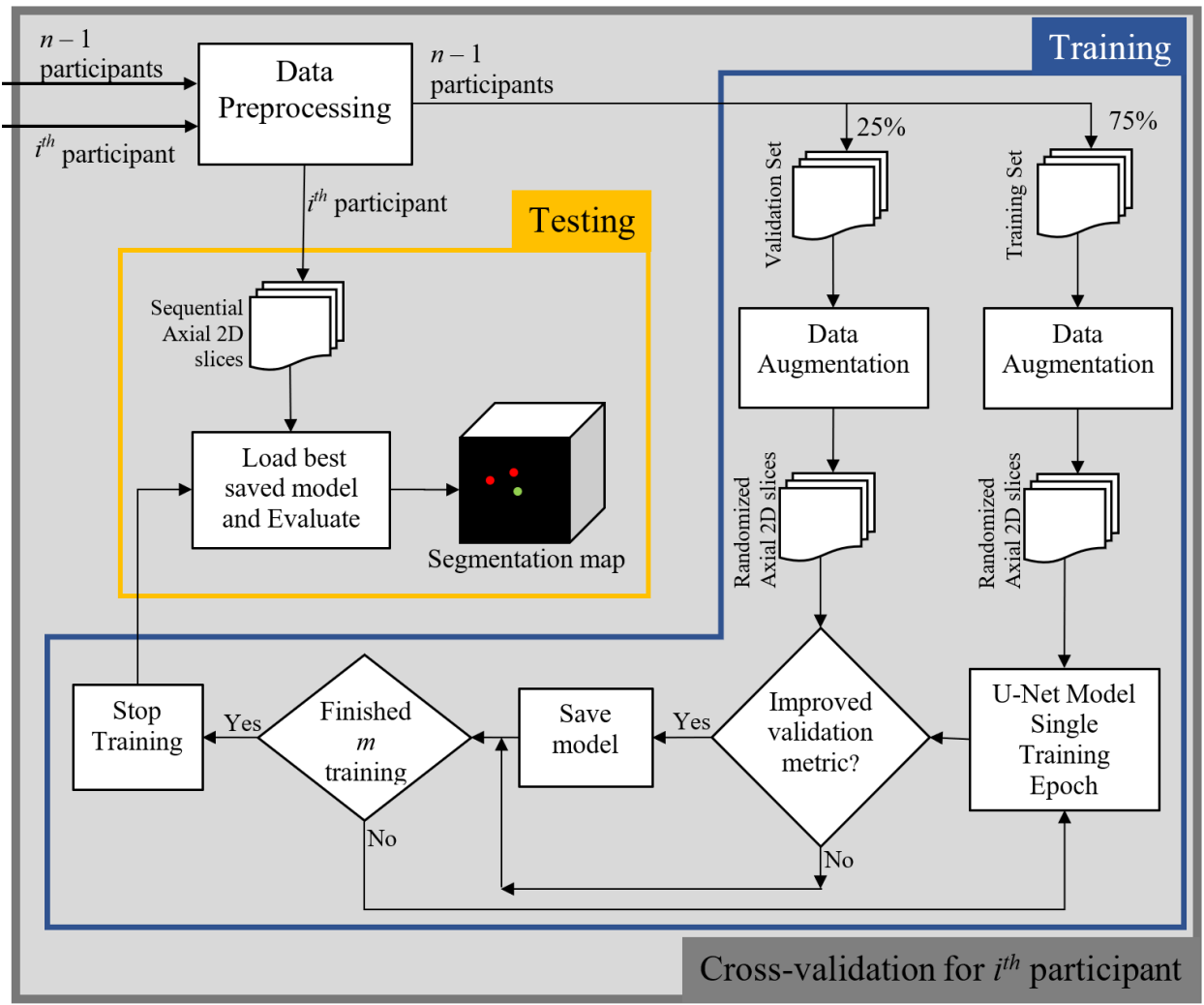

**Figure 5.** Overview of the split for one-fold of the cross-validation process that is repeated n times. In each fold, the model that was used to predict the test participant was trained on the remaining n-1 samples in order to avoid data leakage. Within the training stage, 25 percent of the n-1 participants were used as the validation set. The model with the highest validation accuracy was chosen to predict the left-out participant sample.

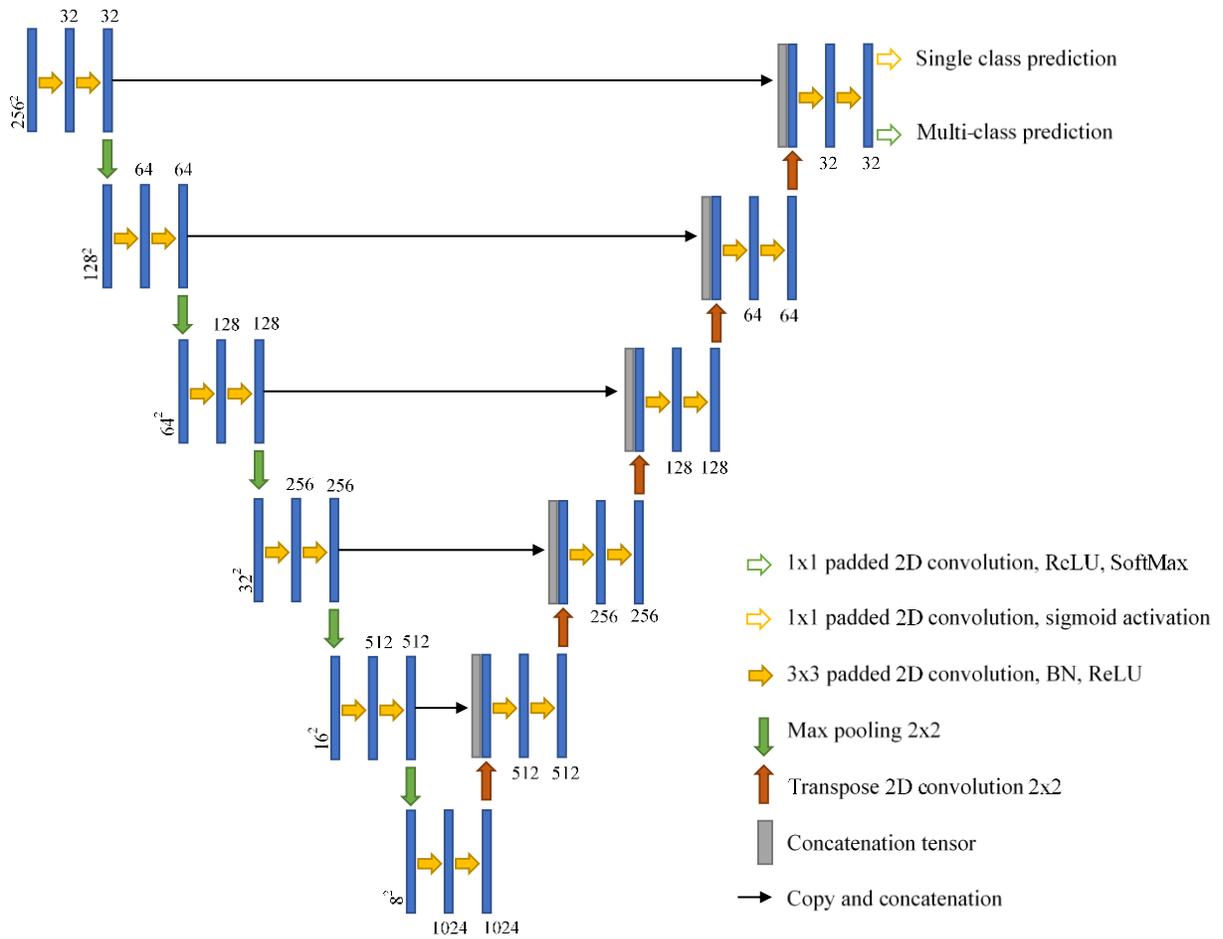

**Figure 6.** U-Net architecture using padded convolutions for both single class and multiclass predictions.

|  | Experiments | Avg Sensitivity ± SEM [CI: lower, upper] | Avg Precision ± SEM [CI: lower, upper] | Avg Magnitude Accuracy ± SEM [CI: lower, upper] | Pearson Correlation Coeff (p-value) | Bland-Altman Plot (md, [lower, upper]) |
|---|---|---|---|---|---|---|
| **Single Class CMB** | SWI | 0.85 ± 0.06 CI: [0.74, 0.97] | 0.22 ± 0.04 CI: [0.14, 0.31] | 0.91 ± 0.06 CI: [0.80, 1.03] | 0.96 (p=0.000) | md=-4.54 CI: [-24.77, 15.69] |
| | SWI and QSM | 0.88 ± 0.06 CI: [0.77, 0.99] | 0.40 ± 0.07 CI: [0.27, 0.54] | 1.09 ± 0.04 CI: [1.00, 1.17] | 0.97 (p=0.000) | md=-2.21 CI: [-19.45, 15.04] |
| | SWI and T2w | 0.84 ± 0.07 CI: [0.70, 0.97] | 0.29 ± 0.06 CI: [0.17, 0.42] | 0.95 ± 0.08 CI: [0.80, 1.10] | 0.76 (p=0.000) | md=-3.04 CI: [-36.59, 30.51] |
| | **SWI, QSM and T2w** | 0.87 ± 0.06 CI: [0.76, 0.98] | 0.50 ± 0.07 CI: [0.35, 0.64] | **1.08 ± 0.07 CI: [0.94, 1.22]** | 0.98 (p=0.000) | md=-0.71 CI: [-15.26, 13.84] |
| **Single Class Iron Deposits** | SWI | 0.81 ± 0.06 CI: [0.68, 0.94] | 0.51 ± 0.07 CI: [0.37, 0.65] | 1.06 ± 0.08 CI: [0.91, 1.21] | 0.88 (p=0.000) | md=-4.08 CI: [-102.63, 94.46] |
| | SWI and QSM | 0.77 ± 0.06 CI: [0.65, 0.89] | 0.60 ± 0.07 CI: [0.46, 0.75] | 1.09 ± 0.05 CI: [0.99, 1.20] | 0.92 (p=0.000) | md=-2.54 CI: [-80.42, 75.34] |
| | SWI and T2w | 0.77 ± 0.06 CI: [0.64, 0.89] | 0.56 ± 0.07 CI: [0.42, 0.70] | **1.04 ± 0.08 CI: [0.88, 1.19]** | 0.85 (p=0.000) | md=9.04 CI: [-91.48, 109.56] |
| | SWI, QSM and T2w | 0.81 ± 0.05 CI: [0.71, 0.92] | 0.62 ± 0.07 CI: [0.47, 0.76] | 1.11 ± 0.05 CI: [1.02, 1.21] | 0.81 (p=0.000) | md=2.58 CI: [-114.33, 119.50] |

SEM = Standard error of the mean
md = mean difference
CI = confidence interval
**Bold – Model with highest magnitude accuracy**

**Table 1.** Experimental results using the single class model for the number of predicted CMB and iron deposit lesions evaluated against the reference annotation.

|  | Experiments | Avg Sensitivity ± SEM [CI: lower, upper] | Avg Precision ± SEM [CI: lower, upper] | Avg Magnitude. Accuracy ± SEM [CI: lower, upper] | Pearson Correlation Coeff (p-value) | Bland-Altman Plot (md, [lower, upper]) |
|---|---|---|---|---|---|---|
| **Multi-class CMB** | SWI | 0.82 ± 0.07 CI: [0.68, 0.96] | 0.36 ± 0.06 CI: [0.24, 0.49] | 0.99 ± 0.06 CI: [0.87, 1.12] | 0.96 (p=0.000) | md=0.04 CI: [-26.61, 26.70] |
|  | SWI and QSM | 0.84 ± 0.07 CI: [0.70, 0.98] | 0.59 ± 0.08 CI: [0.43, 0.75] | 1.15 ± 0.07 CI: [1.00, 1.29] | 0.99 (p=0.000) | md=0.67 CI: [-18.26, 19.59] |
|  | SWI and T2w | 0.76 ± 0.08 CI: [0.60, 0.91] | 0.43 ± 0.06 CI: [0.31, 0.56] | 1.00 ± 0.07 CI: [0.86, 1.13] | 0.97 (p=0.000) | md=0.75 CI: [-25.47, 26.97] |
|  | SWI, QSM and T2w | 0.89 ± 0.05 CI: [0.79, 1.00] | 0.49 ± 0.06 CI: [0.37, 0.61] | 1.07 ± 0.06 CI: [0.95, 1.19] | 0.98 (p=0.000) | md=0.92 CI: [-25.49, 27.33] |
| **Multi-class Iron Deposits** | SWI | 0.76 ± 0.06 CI: [0.63, 0.88] | 0.70 ± 0.08 CI: [0.55, 0.86] | 1.13 ± 0.07 CI: [0.99, 1.28] | 0.91 (p=0.000) | md=13.83 CI: [-62.08, 89.75] |
|  | SWI and QSM | 0.75 ± 0.07 CI: [0.62, 0.88] | 0.75 ± 0.08 CI: [0.60, 0.91] | 1.20 ± 0.05 CI: [1.11, 1.30] | 0.91 (p=0.000) | md=6.71 CI: [-74.77, 88.18] |
|  | SWI and T2w | 0.76 ± 0.06 CI: [0.64, 0.89] | 0.60 ± 0.08 CI: [0.44, 0.75] | 1.08 ± 0.07 CI: [0.93, 1.22] | 0.87 (p=0.000) | md=18.29 CI: [-71.54, 108.12] |
|  | SWI, QSM and T2w | 0.81 ± 0.05 CI: [0.71, 0.92] | 0.64 ± 0.08 CI: [0.49, 0.79] | 1.17 ± 0.05 CI: [1.08, 1.27] | 0.90 (p=0.000) | md=16.29 CI: [-62.23, 94.82] |

SEM = Standard error of the mean
md = mean difference
CI = confidence interval
**Bold – Model with highest magnitude accuracy**

**Table 2**. Experimental result using the multiclass model for the number of predicted CMB and iron deposit lesions evaluated against the reference annotation.